\documentstyle[11pt,paspconf,galdyn,psfig]{article}

\begin{document}
\title{Dynamical Constraints on Disk Masses}
\author{J. A. Sellwood}
\affil{Department of Physics \& Astronomy, Rutgers University, \\
136 Frelinghuysen Road, Piscataway, NJ 08854, USA\\
email: sellwood@astro.rutgers.edu}

\begin{abstract}
While the total interior mass of a galaxy is reasonably well determined by a 
good rotation curve, the relative contributions from disk, bulge and halo are 
only weakly constrained by one-dimensional data.  Barred galaxies are 
intrinsically more complicated, but provide much tighter constraints on the disk 
masses and support the idea that most of the mass in the inner parts of bright 
galaxies is in their stars.  There appears to be no systematic difference in 
dark matter content between barred and unbarred galaxies, consistent with the 
theoretical result that the global stability of galaxies with dense centers does 
not depend on their halo fraction.  The rotation curve shapes of lower 
luminosity and low-surface-brightness galaxies, on the other hand, indicate 
significant mass in the DM halo even near their centers.  We find that most DM 
halos appear to have large cores, inconsistent with the predictions from 
cosmological simulations.  We also show that such large-core halos can result 
from compression by disk infall of physically reasonable initial halos.  Maximum 
disks, while apparently required by the data, do seem to present some puzzles; 
most notably they re-open the old disk-halo ``conspiracy'' issue and incorrectly 
predict that surface brightness should be a second parameter in the Tully-Fisher 
relation.
\end{abstract}

\section{Surface brightness and luminosity}
The question of what fraction of the central attraction should be attributed to 
dark matter (DM) within the disk of a spiral galaxy is still unresolved.  Most 
of the controversy surrounds the higher luminosity, high surface brightness 
(HSB) galaxies which I argue here have very little DM in their inner regions.  
As the rotation curves of most LSB and low-luminosity galaxies, on the other 
hand, do not have the shapes predicted from their light distributions, a 
significant DM contribution is required in their inner parts.  While this 
conclusion has been established from much careful work on individual galaxies, 
trends suggesting increasing DM content towards both types of galaxy can also be 
found in a statistical analysis of a large galaxy sample.

\begin{figure}
\centerline{\psfig{figure=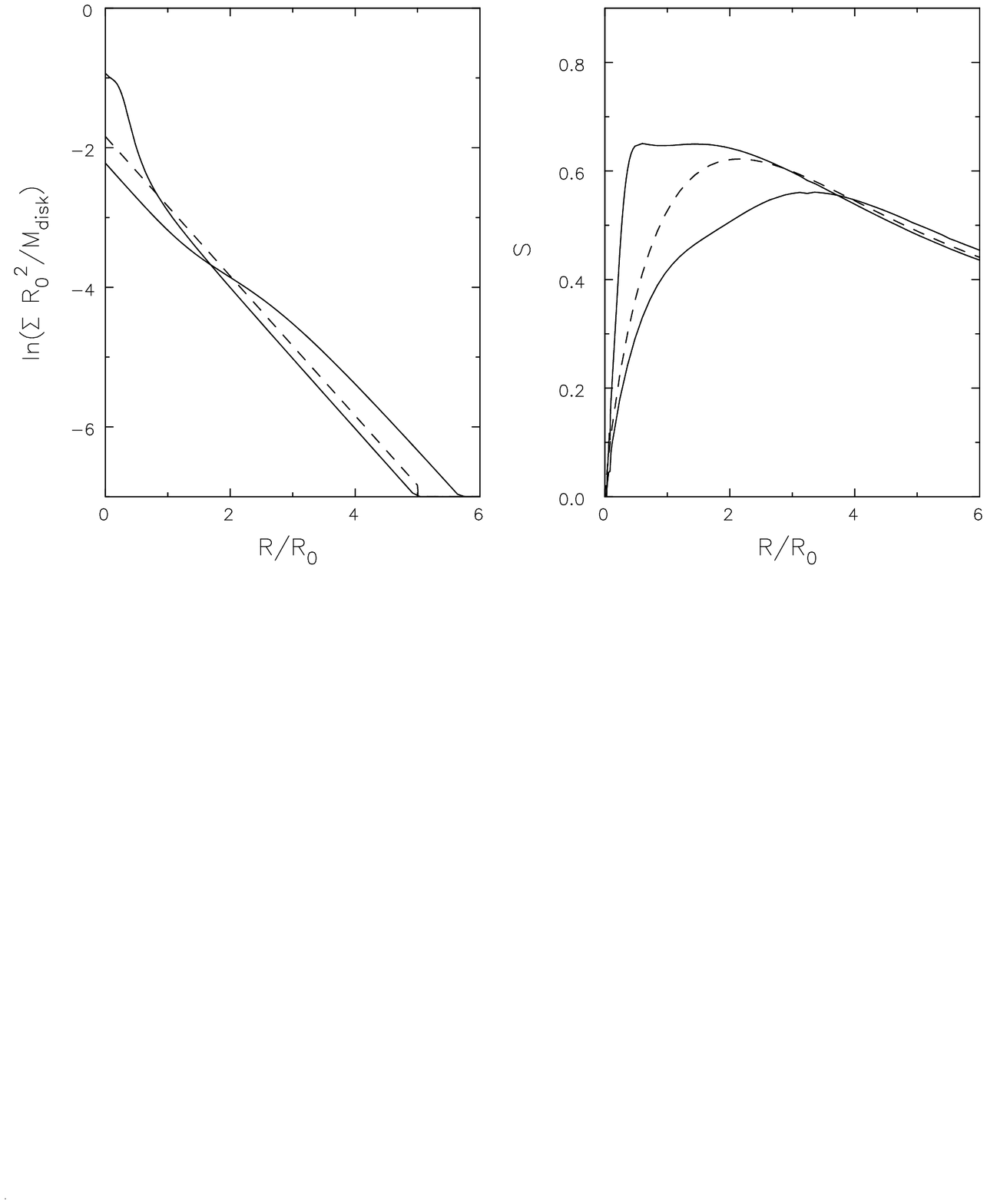,width=0.9\hsize,angle=0}}
\caption{Surface density profiles (left) and corresponding rotation curves 
(right) arising from three different thin disks of equal mass and outer scale 
length.} \label{fig:shapes}
\end{figure}

If the surface density of a disk of mass $M$ decreases exponentially in its 
outer parts with scale length $R_0$, the rotation speed arising from the disk 
alone may be written
\be
V_{\rm disk}(R) = \sqrt{{GM_{\rm disk} \over R_0}} \; S\left({R \over 
R_0}\right)
\ee
The (dimensionless) function $S$, which describes the shape, is plotted in 
Figure \ref{fig:shapes} for three different disks having the same total mass and 
outer exponential scale length.  It can be seen that while the shape of the 
rotation curve arising from the disk is strongly dependent on the surface 
density profile of the inner disk, the height of the maximum is not; it varies 
from 0.6 by about 10\% only between the three cases.  Thus the peak of the 
disk's contribution to the rotation speed in most galaxies $V_{\rm disk,max} 
\sim 0.6\sqrt{GM_{\rm disk} / R_0}$ even when the inner surface density profile 
departs significantly from exponential.

\begin{figure}
\centerline{\psfig{figure=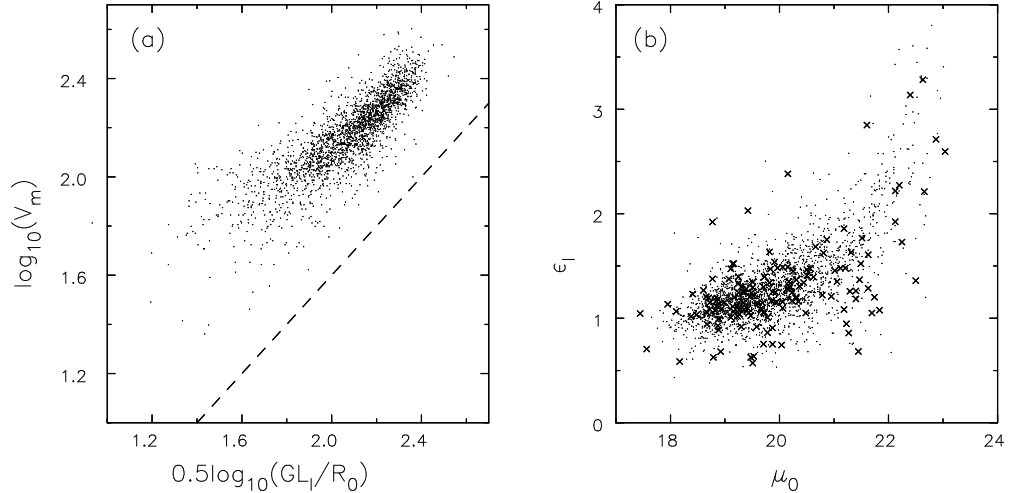,width=\hsize,angle=0}}
\caption{Evidence for increasing DM fractions in both (a) low-luminosity and (b) 
LSB galaxies.  Galaxies classified as barred are marked with crosses in (b).} 
\label{fig:syer}
\end{figure}

We can use this fact to rank galaxies according to the relative contributions of 
the disk and halo to the observed peak rotation speed, $V_{\rm m}$.  Following 
Syer, Mao \& Mo (1998), we form the ratio of observable quantities
\be
\epsilon_{\rm I} = {V_{\rm m} \over \sqrt{GL_{\rm I}/R_0}} {1 \over \sqrt h} 
\qquad \sim 0.6 {V_{\rm m} \over V_{\rm disk,max}} \sqrt{{\Upsilon_{\rm I}\over 
h}}, \label{eq:syer}
\ee
where $L_{\rm I}$ is the disk luminosity (in Solar units) in the I-band.  The 
second form shows that the values obtained depend both on the disk M/L, 
$\Upsilon_{\rm I}$, and on $h \; (= H_0/100$ km s\per\ Mpc\per), because both 
luminosity and scale length are distance dependent.  More than 2000 galaxies 
from the sample gathered by Mathewson \& Ford (1996) are plotted in Figure 
\ref{fig:syer}; galaxies with $V_{\rm sys}<1000\;$km~\per\ are omitted.  
Combining the information they provide with the {\it assumption\/} that the disk 
is exponential, allows us to deduce $R_0$ and the central surface brightness, 
$\mu_0$, in the manner described by Syer \etal

The dashed line in Figure \ref{fig:syer}(a) indicates the slope of simple 
proportionality between $V_{\rm m}$ and $\sqrt{GL_{\rm I}/R_0}$; it is clear 
that the distribution of points towards the low $V_{\rm m}$ (low luminosity) end 
has a shallower slope, indicating increasing DM fractions in these galaxies.  
The right hand panel shows a trend of increasing DM content from high to low SB 
galaxies.  Both trends are strong enough to show through the scatter which is 
increased by deviations from exponential, bulge contributions, intrinsic M/L 
variations, departures from Hubble flow, imperfect inclination and extinction 
corrections, \etc

The trends seen in Figure \ref{fig:syer} suggest increasing mass discrepancies 
within the optical disks towards both low-luminosity and LSB galaxies.  But 
because M/L is so hard to pin down, such diagrams do not tell us whether the 
disks in galaxies having small values of $\epsilon_{\rm I}$ are maximal.  [A 
``maximum disk'' model is one in which $V_{\rm disk,max} / V_{\rm m} \gtsim 
0.85$ (\eg\ Sackett 1997).  This ratio cannot be much larger than 0.85 when a 
model includes an extended DM halo with a density profile that decreases 
monotonically with radius.]

Disagreements arise over how much DM to add to the disk contribution in order to 
fit the observed rotation curve.  As pointed out by van Albada \etal\ (1985; see 
also Navarro 1998), the appropriate M/L for the disk is not constrained at all 
by conventional goodness-of-fit estimators, such as $\chi^2$, especially since 
the mass profile of the DM is unknown.  Here I review some arguments that bear 
on the masses of disks and refer the reader to others given by Bosma elsewhere 
in these proceedings.  The DM content of LSB galaxies is discussed by de Blok 
(also this volume).

\section{No halo fits}
The constraint on the disk M/L is especially weak when low spatial resolution 
21cm data are used to determine the inner rotation curve; tighter constraints 
are furnished by optical data.  Many authors (Kalnajs 1983; Kent 1986; Buchhorn 
1992; Broeils \& Courteau 1997; Palunas \& Williams 1999) have successfully 
fitted pure disk models to optical rotation curve data.  In this approach, one 
assumes a constant M/L for the disk, and sometimes a separate value for the 
bulge, and determines the central attraction from an exact solution of Poisson's 
equation for the mass distribution thus inferred from the light.  The shape of 
the rotation curve predicted in this manner generally bears a remarkably close 
resemblance to that observed, at least in the inner disk.

\begin{figure}[t]
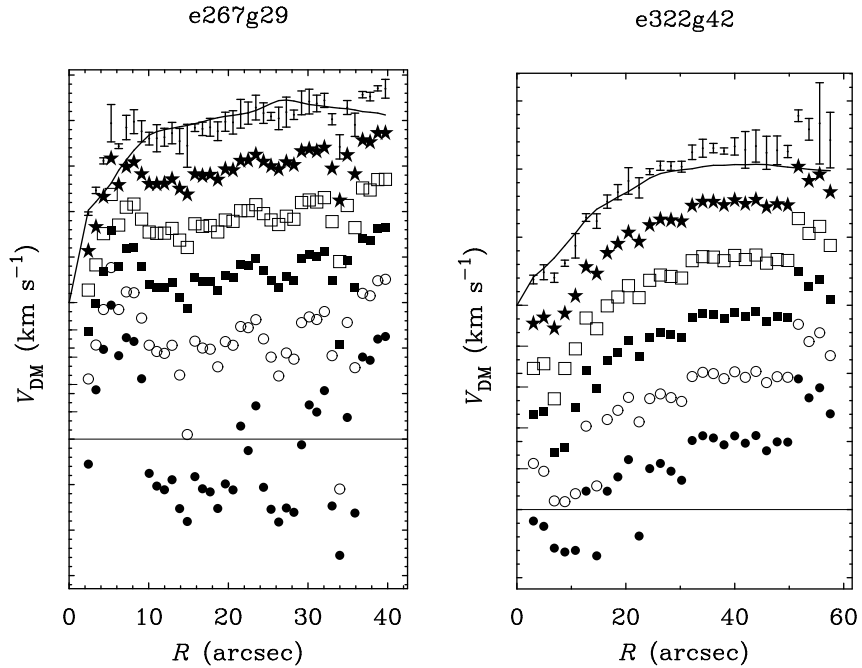

\centerline{\psfig{figure=sellwood_3a.ps,width=0.4\hsize,angle=0}
\hskip0.5cm \psfig{figure=sellwood_3b.ps,width=0.4\hsize,angle=0}}
\caption{Halo rotation curves required for sub-maximum disks.  The error bars 
show the measured circular speed and the curve shows the pure disk fit (the 
bulge contribution is omitted in e267g29).  The symbols show the halo 
contribution required if the disk has 100\% (filled circles), 80\% (open 
circles), 60\% (filled squares), 40\% (open squares) and 20\% (stars) of the 
maximum value.  The horizontal line indicates the vertical zero point for the 
halo velocities required for 100\% disk.  Subsequent curves are shifted upwards 
by 30~km~s\per\ for every 20\% decrease in disk mass.   Note the similarity in 
shapes of the curves marked by symbols and the error bars.} \label{fig:submax}
\end{figure}

The largest possible M/L value allowed by the observed circular speed in the 
inner disk leads to an impressive fit, with discrepancies noticeable near the 
outer edge only in some cases.  The ``bumps and wiggles'' seen in rotation 
curves from single-slit observations have long been thought to arise from spiral 
arm streaming (see also Bosma, this volume).  This suspicion has been confirmed 
in the 2-D velocity maps obtained by Palunas \& Williams (1999) which allow most 
non-axisymmetric flows to be removed.

The principal conclusion from previous work survives, namely that the overall 
shape of the rotation curve is well predicted by the light distribution.  As 
previously remarked by van Albada \& Sancisi (1986), Freeman (1992) and others,
reducing the M/L for the disk would require DM mass distributions tailored 
individually to match the rotation curve shape of each galaxy.  Two illustrative 
examples taken from the Palunas \& Williams sample are shown in Figure 
\ref{fig:submax}.

\begin{figure}
\centerline{\psfig{figure=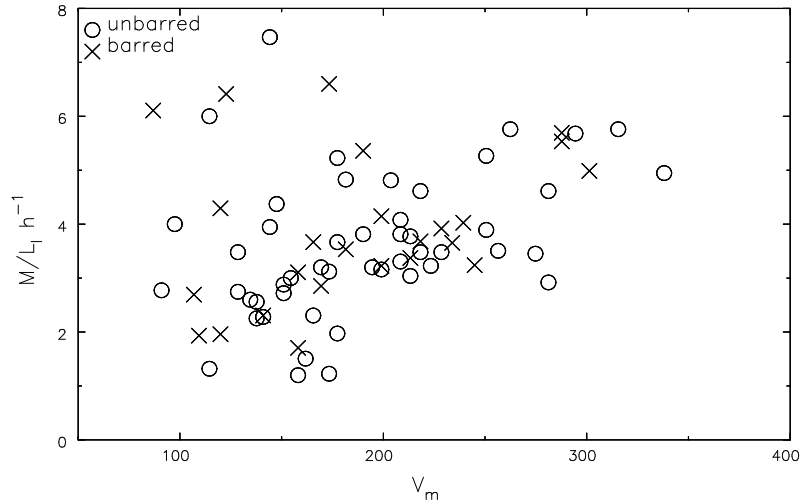,width=0.8\hsize,angle=0}}
\caption{M/L$_{\rm I}$ values for maximum disks in the Palunas \& Williams 
sample.  The values plotted should be reduced by a factor $h$.} \label{fig:mtol}
\end{figure}

Figure \ref{fig:mtol} shows the distribution of disk M/L values obtained by 
Palunas \& Williams for their no-halo fits.  One can see a smaller spread of 
values for high luminosity galaxies ($V_{\rm m} > 200\;$km/s), with perhaps a 
hint of a higher M/L in earlier Hubble types.  The broader spread for the lower 
luminosity galaxies could be due to two factors; first, it is harder to 
determine the inclination for these often strongly non-axisymmetric galaxies and 
second, there are some galaxies for which large M/L values give acceptable fits, 
even though one might expect substantial DM fractions in these low-luminosity 
systems.  It is also worth noting that the M/L$_{\rm I}$ values obtained for the 
luminous galaxies (for $h \sim 0.6$) are in line with values predicted by 
Jablonka \& Arimoto (1992) and by Worthey (1994) for quite reasonable population 
models.

As already noted, these ``no halo'' fits frequently reveal a mass discrepancy 
near the edge of the optical rotation curve.  Conventional ``maximum disk'' 
models require a halo with a large, low-density core (\eg\ Broeils 1992).  
Figure \ref{fig:submax} shows that fits with a non-hollow halo would require at 
most a slight reduction in the disk M/L.

None of these arguments is either new or truly compelling.  Others (\eg\ van der 
Kruit 1995; Navarro 1998; Courteau \& Rix 1999) stress that less than maximum 
disk models also yield acceptable fits.  It has been argued (\eg\ Blumenthal 
\etal\ 1986) that halo compression as the disk forms leads to featureless 
rotation curves (see also \S6), and this argument is extended somehow to account 
for the similarity of the shape of halo contribution to that of the disk.  The 
population synthesis argument is weak because minor changes to the low-mass end 
of the IMF can lead to significant changes in the predicted M/L.

\section{Barred galaxy velocity fields}
The above argument is not decisive because we have considerable freedom to 
decompose a 1-D rotation curve into contributions from the sum of two, or more, 
other 1-D functions.  Our group at Rutgers has therefore embarked on a program 
to use the 2-D velocity fields of barred spiral galaxies to provide extra 
constraints.

A massive bar in a disk galaxy distorts the usual circular flow pattern, leading 
to characteristic `$\cal S$'-shaped iso-velocity contours when such a galaxy is 
observed in a suitable projection.  The strength of the non-axisymmetric flow 
pattern can be modeled to determine the mass required in the barred component of 
the potential, leading to an estimate of the disk M/L that is independent of 
rotation curve fitting.

The first such study has been completed for the southern barred spiral galaxy 
NGC~4123.  Weiner (1998) has collected broad-band photometry and velocity maps 
both for the inner galaxy, using Fabry-Perot measurements of the H$\alpha$ 
emission, and for the outer galaxy, using 21~cm data from the VLA.  He has 
constructed models based on full 2-D hydrodynamical simulations of a massless 
gas layer in a rotating potential derived from the observed light  distribution.

In order to construct the model potential from his I-band image, Weiner 
subtracts an unresolved source at the center, rectifies the image to face on, 
assumes a constant thickness and computes the gravitational field, modulo a 
single unknown M/L.  For a number of assumed M/L values, he constructs 
axisymmetric DM halos which, when combined with the disk contribution, fit the 
rotation curve well outside the barred region.

He has run a grid of models covering the parameter space of two unknowns: the 
M/L and the pattern speed of the bar and compares the projected velocity 
distribution with the high spatial resolution Fabry-Perot velocity field in the 
inner parts of NGC~4123.  He finds (with a very high degree of confidence) $2.6 
\leq {\rm M/L_I} h^{-1} \leq 3.3$ is required to produce a flow pattern with 
strong enough non-circular motions to match the data.

This range of M/L values requires that 72\% to 90\% of the mass within 10 kpc is 
in the stars -- the contribution from DM can be little more than the minimal 
amount required to prevent a hollow halo.  Work on other galaxies is in hand to 
check that this is not just a freak case.  It should be noted that Englmaier \& 
Gerhard (this volume) reach a similar conclusion from similar work on the Milky 
Way.

\section{Dynamical friction on bars}
A second argument from barred galaxies relates to the prediction (Weinberg 1985) 
that bars should be slowed dramatically by dynamical friction.  This prediction 
has been confirmed (Debattista \& Sellwood 1998) for moderate halo masses in 
$N$-body simulations of a barred disk in a responsive halo.  The strong 
retarding torque acting on the bar causes the pattern speed to decrease rapidly 
to about one fifth of its initial value.  Since this change occurs while the bar 
length increases only marginally, the ratio, $\cal R$, of the corotation radius 
to the bar semi-major axis, increases from $\gtsim 1$ at a time soon after the 
bar formed, to some ${\cal R} \sim 2.5$ late in the simulation.

On the other hand, $\cal R$ remains close to 1.3 in a ``maximum disk'' model 
having a realistic rotation curve for as long as the calculation was run.

Further work (Debattista \& Sellwood, in preparation) has shown that friction is 
only moderately reduced in halos that rotate in the same sense as the disk, even 
when halo rotation is cranked up to a perhaps unrealistic extent.  Friction is 
also largely independent of whether the velocity distribution of a non-rotating 
halo is isotropic, radially or azimuthally biased, but friction was reduced when 
an anisotropic halo was given a high degree of rotation.

The position of corotation is not easily determined in real barred galaxies.  
Weiner (\S3) finds $1 \ltsim {\cal R} \ltsim 1.4$ for NGC~4123.  A value ${\cal 
R} \simeq 1.2$ was deduced by Lindblad \etal\ (1996) from a very similar study 
of NGC~1365.  More direct estimates can be made for SB0 galaxies using the 
technique proposed by Tremaine \& Weinberg (1984) which requires that the 
observed material, stars in this case, obeys an equation of continuity.  
Application of this method to NGC~936 by Merrifield \& Kuijken (1995) and to 
NGC~4596 by Gersson \etal\ (this volume) also places corotation at a radius 
$\ltsim 1.5$ times the bar semi-major axis.  The shapes and locations of dust 
lanes in many other barred galaxies also suggest a ratio of 1.2 (Athanassoula 
1992) and finally some still more model-dependent studies of ringed galaxies 
(Buta \& Combes 1995) suggest a similar value.

While this heterogeneous collection of estimates is not as solid as one would 
like, all evidence is consistent with ${\cal R} \ltsim 1.5$, implying that real 
bars have not experienced strong braking.  Once again, therefore, the DM halo 
must have a large, low-density core if the radius of corotation is to stay as 
close to the bar end as observations seem to imply.

\begin{figure}
\psfig{figure=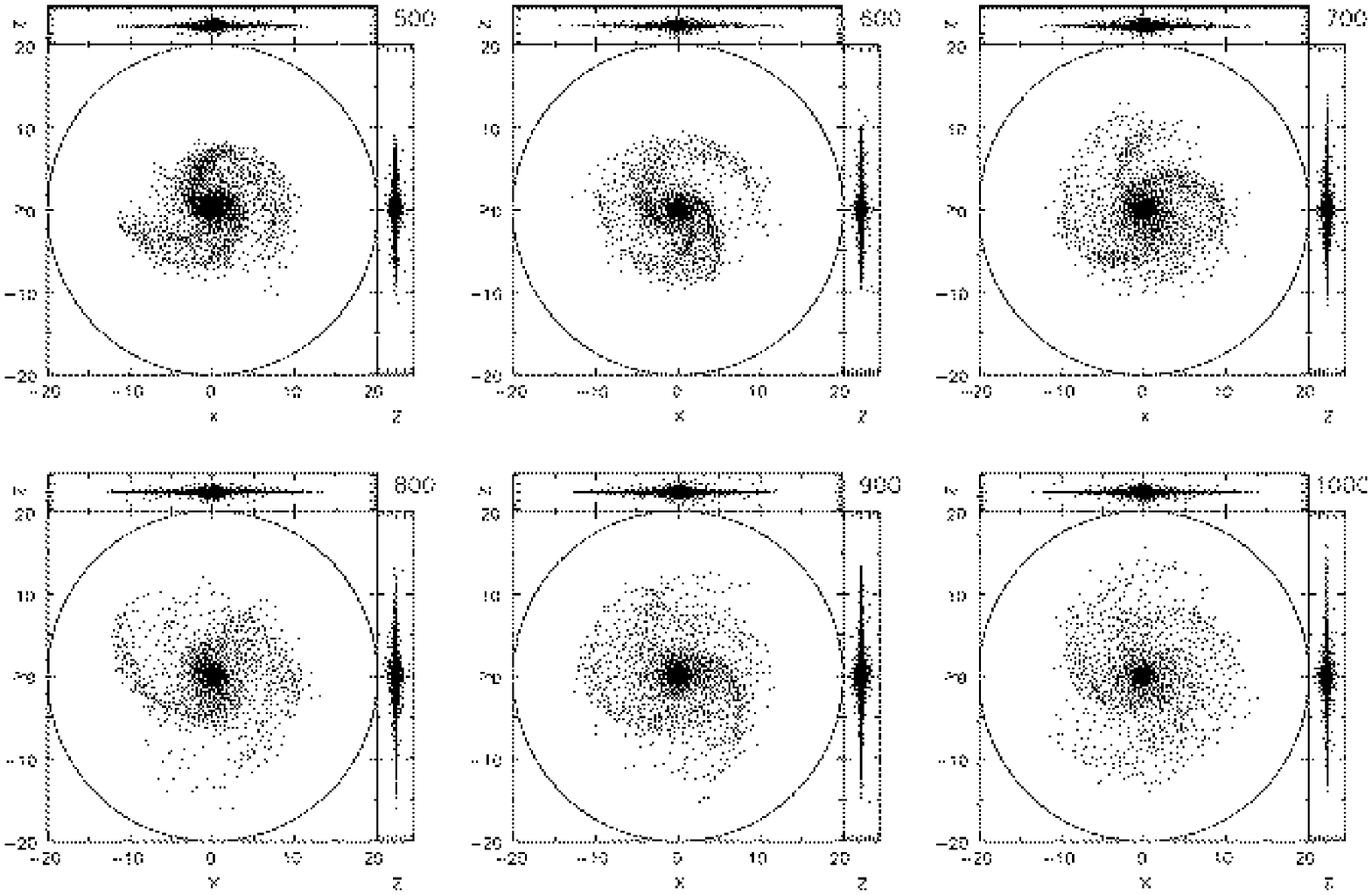,width=\hsize,angle=0}
\psfig{figure=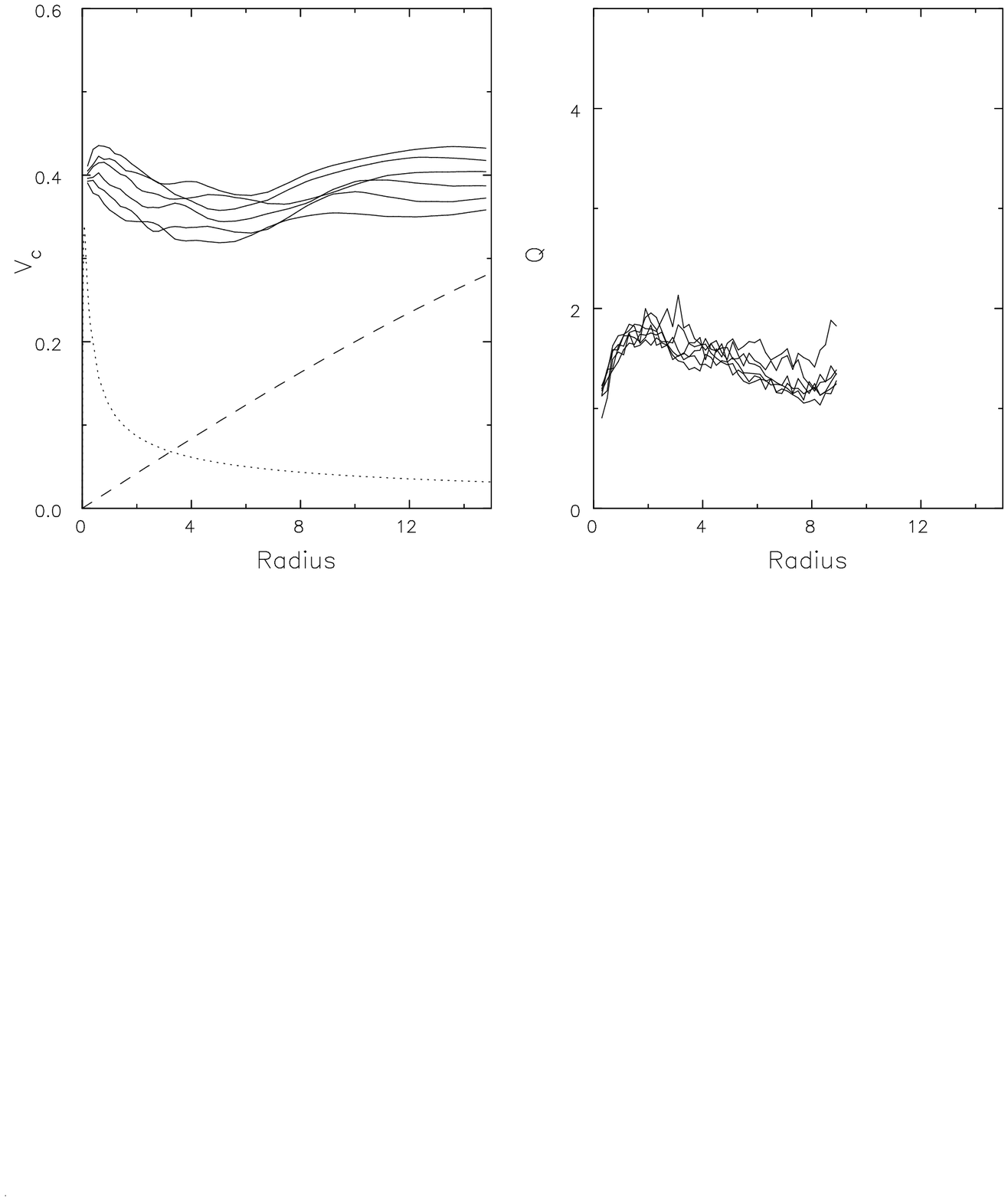,width=\hsize,angle=0}
\caption{Simulation of a massive, cool and bar-stable disk.  The top panels show 
the second half of the evolution in which strong spiral patterns are present but 
no bar.  The rotation curves (below left) are measured at the times illustrated 
and show the gradual increase in disk mass; the fixed contributions from the 
rigid halo (dashed) and central mass (dotted) are shown.  The values of $Q$ are 
also plotted (below right).} \label{fig:stable}
\end{figure}

\section{Stability}
Since local stability arguments bearing on the question of the appropriate disk 
mass are used by Fuchs (this volume) and his arguments are critically reviewed 
by Bosma (also this volume), I avoid repeating them here and confine this part 
of my discussion to the question of global bar stability only.

It has been known ever since the pioneering $N$-body simulations by Miller 
\etal\ (1971) and Hohl (1971) that self-gravitating galaxy disk models are prone 
to a bar-forming instability.  In a widely cited paper, Ostriker \& Peebles 
(1973) argued that the global stability of unbarred galaxies required a 
significant DM content which, as stressed by Kalnajs (1987), must reside in the 
inner galaxy.  This idea is too simple, however.  Toomre (1981) argued that 
galaxies can be stabilized by other means; his argument is restated by Binney \& 
Tremaine (1987, \S6.3).

The fact that $m=2$ modes in an almost fully self-gravitating disk can be 
stabilized by a dense center has been shown in linear stability analyses by Zang 
(1976), Evans \& Read (1998) and Toomre (unpublished).  I was able (Sellwood 
1989) to confirm Toomre's linear theory predictions, but I found that the 
stability of the extreme model he chose was rather delicate, since bars were 
still triggered by quite modest finite-amplitude effects.  These extreme cases 
also suffer from $m=1$ instabilities.

Not all models stabilized in this way are delicate, however.  Figure 
\ref{fig:stable} shows a model that is robustly stable to bar formation, even 
though it has a moderately cool massive disk in a diffuse halo.  This model 
closely resembles those described by Sellwood \& Moore (1999), but the radial 
distribution of the added particles is different.  By concentrating all the 
infalling particles into an rather narrow annulus, Sellwood \& Moore were able 
to show that spiral patterns could be strong enough to re-arrange the surface 
density and alter the rotation curve shape.  Since the strong spirals that 
achieved this result led to a rather hot disk ($Q \sim 4$), I added particles in 
the model presented here over a wide radial range, which allowed the disk to 
remain cool as shown.

Almost all the central attraction in this model comes from the disk; it supports 
strong two-arm spiral patterns yet does not form a bar.  The key difference 
between models of this kind and previous bar-unstable disks is the steep inner 
rise of the rotation curve, caused mostly by mobile particles, but seeded by the 
introduction of a fixed mass, having $\sim1$\% of the final disk mass.

The steep central rise in the rotation curve of this model resembles that of 
many galaxies of both late (Rubin, Ford \& Thonnard 1980) and early (Rubin, 
Kenney \& Young 1997) Hubble types.

The conclusion of this section is that considerations of global bar-stability do 
not require a high central density of DM in every galaxy.  There is no universal 
stability criterion; as Ostriker \& Peebles argued, galaxies with gently rising 
curves are unstable unless the disk is significantly sub-maximum, but we now 
know that fully self-gravitating disks with steeply rising curves are stable.

Strong evidence that the existence or absence of a bar has nothing whatsoever to 
do with DM content can be seen in Figures \ref{fig:syer}(b) and \ref{fig:mtol} 
in which the barred and unbarred galaxies are marked by different symbols.  
There is no apparent tendency for barred galaxies to have lower $\epsilon_{\rm 
I}$ or higher M/L than their unbarred counterparts.

\section{Halo compression}
As the principal source of central attraction switches from disk to halo matter, 
we might expect the rotation curve of a maximum disk galaxy to have a feature 
near the disk edge.  Bahcall \& Casertano (1985), among others, stressed the 
absence of such a feature, which became known as the ``disk-halo conspiracy.''  
In fact, it is a serious overstatement to suggest that there is no such feature 
-- quite a number of galaxies are known in which a marked decrease in circular 
speed is observed near the disk edge; Bosma (this volume) gives a list of some 
of the well-established cases and other examples can be found (\eg\ Verheijen 
1997).  Nevertheless, a weaker conspiracy remains because the circular speed 
well beyond the optical disk is very rarely less than 90\% of that in the disk 
region.

Amongst enthusiasts for dynamically significant DM even near the centers of 
galaxies, the flatness of galaxy rotation curves is regarded as a natural 
consequence of halo compression as the disk forms within it (Blumenthal \etal\ 
1986; Flores \etal\ 1993).  If disks are maximum, however, the close 
correspondence between the circular speeds in the disk and halo still requires a 
conspiracy.

Navarro (1998) has taken the argument further, to suggest that halos with large 
cores are unphysical because the standard formula for halo compression implies a 
hollow, or even negative density, halo before the disk formed.  The problem 
discovered by Navarro, however, is not an argument against large cores, but 
merely reveals the severe limitations of the standard halo compression formula 
(hereafter HCF).

\begin{figure}[t]
\centerline{\psfig{figure=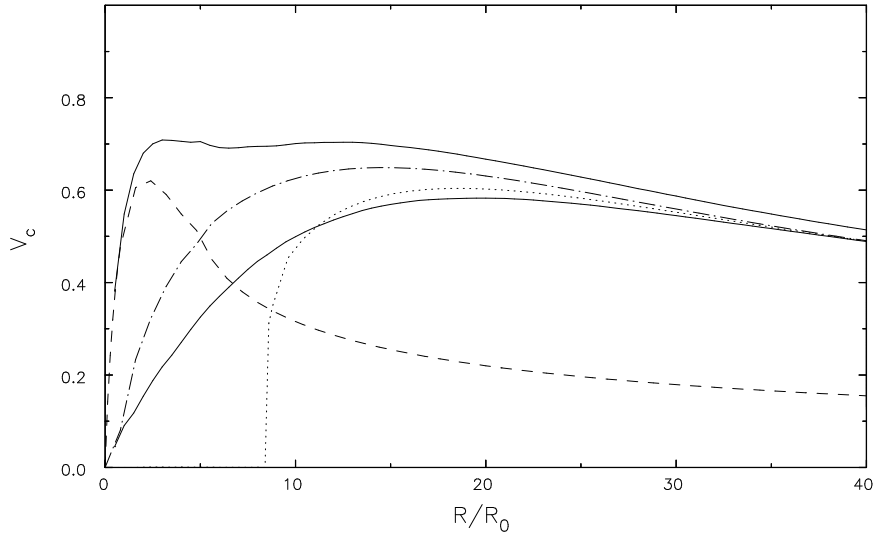,width=0.9\hsize,angle=0}}
\caption{Compression of a spherical halo by a maximum disk.  The solid curve 
which peaks near $R=20R_0$ is the rotation curve of the halo before the disk was 
added.  The upper solid curve shows the combined rotation curve of the disk and 
halo, after adding the disk, and the dot-dashed and dashed curves show the 
contributions of the halo and disk respectively.  The dotted curve shows the 
prediction of the initial halo rotation curve required by HCF.} 
\label{fig:decomp}
\end{figure}

The HCF, as derived by Barnes \& White (1984), Blumenthal \etal\ (1986) and 
Ryden \& Gunn (1987), embodies three principal assumptions.  (1) The mass 
distributions of both the halo and the disk(!)\ are spherical.  (2) The disk 
forms slowly enough that the halo response is adiabatic.  Assumptions (1) and 
(2) imply that halo particles conserve their actions, in particular, the 
component of their angular momentum normal to the orbit plane, $J_z$, 
throughout.  (3) The mean radius of a halo particle's orbit can be computed from 
its ``home radius'' -- the radius of a circular orbit of the same $J_z$.  This 
circular orbit assumption is sometimes stated that shells of halo matter do not 
cross.  These drastic approximations do lead to a single (implicit) relation 
relating the radial mass profiles of the halo before and after the disk formed.  
Barnes (1987) conducted a direct test and concluded that the HCF ``overestimates 
the halo response by as much as a factor of two.''  Despite his warning, the 
na\"\i ve HCF is still widely used.

Figure \ref{fig:decomp} demonstrates its failure for a maximum disk case.  The 
plot shows the formation of a maximum disk galaxy model as an exponential disk 
builds up within an initially spherical, low-concentration halo.  The halo is 
composed of $100\,000$ particles that move freely in their own gravitational 
well supplemented by that of the disk.  The final disk has a total mass of one 
tenth that of the halo and, to avoid any angular momentum redistribution, I held 
the disk particles fixed once they were placed in position.  The disk was grown 
gradually by adding two particles per time-step.  The resulting maximum disk 
model has a flat rotation curve out to $\sim15R_0$.

The dotted curve shows the rotation curve of the pre-compressed halo that would 
be required by the HCF to yield the final maximum disk model.  The fact that 
velocities become imaginary at radii $\ltsim 8R_0$ confirms Navarro's finding 
that the na\"\i ve HCF predicts nonsense.  That the HCF fails badly in this case 
is evidenced by the reasonable initial halo actually required for the final 
maximum disk model.

I conducted further experiments to determine which approximation above is most 
responsible for this gross error.  The simulation shown in Figure 
\ref{fig:decomp} was, in fact, run backwards; I started with an equilibrium 
disk-halo model and evaporated the disk to find the required ``initial'' halo.  
As a check of the adiabatic assumption (2), I re-grew the same disk and 
recovered the rotation curve from which I had started to impressive precision.  
In order to test assumption (1), I ran a new calculation with the mass of the 
flat disk replaced by the equivalent, but now spherical, $M(R)$ profile.  This 
change did alter the rotation curve of the halo after evaporation of the 
``spherical disk'' but still it was far from correctly predicted by the HCF, 
implying that assumption (3) is the principal source of error in the simple 
formula.

While this experiment disposes of the objection that a compressed halo having a 
large core is unphysical, it also highlights the disk-halo conspiracy.  In order 
to yield a reasonably flat rotation curve after the disk has formed, a $V_{\rm 
max}$ for the initial halo similar to that for the disk which will form within 
it seems to be required.  It is far from obvious that such well-tuned initial 
conditions would arise naturally.

\section{Surface Brightness and the Tully-Fisher Relation}
The rotation curve of any galaxy in which the DM contribution in the inner disk 
is negligible, will have the form given in equation (1).  If disks are maximal, 
equation (1) predicts that the observed circular speed should vary as 
$R_0^{-1/2}$ at fixed disk mass, or luminosity if the M/L is approximately 
constant.  Thus one might expect that surface brightness variations to give rise 
to scatter about the Tully-Fisher relation (TFR).  This prediction is well known 
to fail for LSB galaxies; one of the principal surprises from studies of LSB 
galaxies is that they (or at least the larger ones) lie on the same TFR as do 
the HSBs (Zwaan \etal\ 1995; Sprayberry \etal\ 1995).  This result is one strand 
of evidence for large mass discrepancies in their inner parts (de Blok, this 
volume).

If all bright HSB galaxies are maximal, on the other hand, the above predicted 
correlation between circular velocity in the inner disk and scale length should 
hold.  Courteau \& Rix (1999) tested for this in a sample of bright galaxies 
with well measured $R_0$ and circular speed at $R=2.2R_0$, but found that the 
residuals from a TF-like relation do not correlate with surface brightness.  
Thus even within a sample of HSB galaxies with tightly controlled properties, 
surface brightness was not a second parameter in the TFR.

As was similarly inferred for the LSBs (de Blok \& McGaugh 1997), their result 
requires either that the disk M/L varies systematically with surface brightness, 
or that the halo contribution picks up by just enough to compensate for a 
decreasing disk contribution, or that Newtonian dynamics breaks down.  The same 
conclusions can be drawn from the trend in Figure \ref{fig:syer}(b), albeit from 
data of lower quality.  Courteau \& Rix favor the second alternative, which 
evidently requires that at least some higher SB galaxies have significant DM 
contributions to their inner rotation speeds.

\section{Conclusions}
It is now well-established that LSB and low luminosity galaxies have larger mass 
discrepancies in their inner parts than do the bright HSB galaxies (Figure 
\ref{fig:syer}).  The controversial question is how small is the DM contribution 
to the inner rotation curves of the larger HSB galaxies?

Studies of barred galaxies, in particular, have yielded strong new evidence 
suggesting that most mass in their inner parts is in stars.  The completely 
independent arguments presented in \S\S3 \& 4 both suggest low upper limits to 
the DM content of inner parts of barred galaxies.  These limits are not so 
extreme as to violate the conventional maximum disk constraint that the DM halo 
density should not decrease towards the center.  Combined with the apparent 
absence of any systematic offsets between barred and unbarred galaxies in 
Figures \ref{fig:syer}(b) and \ref{fig:mtol}, it seems reasonable to argue that 
all HSB galaxies are similarly deficient in DM in their inner parts.  This 
conclusion is supported by the older evidence from rotation curve fitting (\S2), 
by some recent evidence for the Milky Way -- especially the result obtained by 
Englmaier \& Gerhard (1999, and this volume).  Bosma (this volume) presents 
further arguments for disk masses ranging up to maximum.

The case for maximum disks in bright HSB galaxies is therefore strong, though 
still not decisive, partly because it leaves at least two serious puzzles:  
First, why is the circular speed from the disk in the inner galaxy generally so 
similar to that from the halo further out?  Second, why does DM gradually become 
more important as the disk surface brightness declines?  Courteau \& Rix (1999) 
argue that even HSB galaxies have, on average, substantial DM fractions in their 
inner parts and that barred galaxies have generally smaller fractions.  This 
last suggestion is, however, inconsistent with the evidence in Figures 
\ref{fig:syer}(b) and \ref{fig:mtol}.

The DM halos of LSB and low luminosity galaxies are well known to have large 
cores with low central densities and the evidence presented here suggests this 
is also true for bright HSBs.  DM halos of this type are quite different from 
those predicted in many simulations of hierarchical clustering in an expanding 
universe (but see Primack, this volume).

\acknowledgments
I thank Stacy McGaugh and Scott Tremaine for helpful conversations and both of 
them, as well as Ben Weiner, for comments on the manuscript.  This work was 
supported by NSF grant AST 96/17088 and NASA LTSA grant NAG 5-6037.


\begin{references}
\footnotesize

\reference 
Athanassoula, E. 1992, \mnras, {\bf 259}, 345 

\reference 
Bahcall, J. N. \& Casertano, S. 1986, \apj, {\bf 308}, 347

\reference 
Barnes, J. 1987, in \FaberNNG\ p.~154

\reference 
Barnes, J. \& White, S. D. M. 1984, \mnras, {\bf 211}, 753 

\reference 
Binney, J. \& Tremaine, S. 1987, \BinneyTremaine

\reference 
Blumenthal, G. R., Faber, S. M., Flores, R. \& Primack, J. R. 1986, \apj, {\bf 
301}, 27 

\reference 
Broeils, A. 1992, \PhD, Groningen University

\reference 
Broeils, A. H. \& Courteau, S. 1997, \PersicSalucci\ p.~74

\reference 
Buchhorn, M. 1992, \PhD\ Australian National University

\reference 
Buta, R. \& Combes, F.\ 1996, \fcp, {\bf 17}, 95 

\reference 
Courteau, S. \& Rix, H-W. 1999, \apj, (to appear) (astro-ph/9707290)

\reference 
Debattista, V. P. \& Sellwood, J. A. 1998, \apjl, {\bf 493}, L5

\reference
de Blok, W. J. G. \& McGaugh, S. S. 1997, \mnras {\bf 290}, 533

\reference
Englmaier, P. \& Gerhard, O. 1999, \mnras, to appear

\reference 
Evans, N. W. \& Read, J. C. A. 1998, \mnras, {\bf 300}, 106 

\reference 
Flores, R., Primack, J. R., Blumenthal, G. R. \& Faber, S. M., 1993, \apj, {\bf 
412}, 443 

\reference 
Freeman, K. C. 1992, in \ThuanBT\ p.~201

\reference 
Hohl, F. 1971, \apj, {\bf 168}, 343 

\reference 
Jablonka, J. \& Arimoto, N., 1994, \aap, {\bf 255}, 63

\reference 
Kalnajs, A. J. 1983, in \Athanassoula\ p.~87

\reference 
Kalnajs A. J. 1987, in \KormendyKnapp\ p.~289

\reference 
Kent, S. M. 1986, \aj, {\bf 91}, 1301

\reference 
Lindblad, P. A. B., Lindblad, P. O. \& Athanassoula, E. 1996, \aap, {\bf 313}, 
65

\reference 
Mathewson, D. S. \& Ford, V. L. 1996, \apjs, {\bf 109}, 97

\reference 
Merrifield, M. R. \& Kuijken, K. 1995, \mnras, {\bf 274}, 933

\reference 
Miller, R. H., Prendergast, K. H. \& Quirk, W. J. 1970, \apj, {\bf 161}, 903 

\reference 
Navarro, J. 1998, \apj, (submitted) astro-ph/9807084

\reference 
Ostriker, J. P. \& Peebles, P. J. E. 1973, \apj, {\bf 186}, 467

\reference 
Palunas, P. \& Williams, T. B. 1999, \aj, (to appear)

\reference 
Rubin, V. C., Ford, W. K. \& Thonnard, N. 1980, \apj, {\bf 238}, 471

\reference 
Rubin, V. C., Kenney, J. D. P. \& Young, J. S. 1997, \aj, {\bf 113}, 1250

\reference 
Ryden, B. S. \& Gunn, J. E. 1987, \apj, {\bf 318}, 15

\reference 
Sackett, P. D. 1997, \apj, {\bf 483}, 103

\reference 
Sellwood, J. A. 1989, \mnras, {\bf 238}, 115

\reference 
Sellwood, J. A. \& Moore, E. M. 1999, \apj, {\bf 510}, 125

\reference 
Sprayberry, D., Bernstein, G. M., Impey, C. D. \& Bothun, G. D. 1995, \apj, {\bf 
438}, 72

\reference 
Syer, D., Mao, S. \& Mo, H. J. 1998, astro-ph/9807077

\reference 
Toomre, A. 1981, in \FallLB\ p.~111

\reference 
Tremaine, S. \& Weinberg, M. D. 1984, \apjl, {\bf 282}, L5

\reference 
van Albada, T. S. \& Sancisi, R. 1986, \ptl, {\bf 320}, 447

\reference 
van Albada, T. S., Bahcall, J. N., Begeman, K. \& Sancisi, R. 1985, \apj, {\bf 
295}, 305

\reference 
van der Kruit, P. C. 1995, in \vdKruitGilmore\ p.~205

\reference 
Verheijen, M. 1997, \PhD, University of Groningen

\reference 
Weinberg, M. D. 1985, \mnras, {\bf 213}, 451 

\reference 
Weiner, B. 1998, \PhD, Rutgers University

\reference 
Worthey, G. 1994, \apj Supp, {\bf 95}, 107

\reference 
Zang, T. A. 1976, \PhD, MIT

\reference 
Zwaan, M. A., van der Hulst, J. M., de Blok, W. J. G., \& McGaugh, S. S. 1995, 
\mnras, {\bf 273}, L35

\end{references}
\end{document}